\input phyzzx
\sequentialequations
\overfullrule=0pt
\tolerance=5000
\nopubblock
\twelvepoint
%\magnification=1200
 
%\line{\hfill }
\line{\hfill PUPT 1625, IASSNS 96/52}
\line{\hfill cond-mat/yymmxxx}
\line{\hfill May 1996}

\titlepage
\title{$2n$ Quasihole States Realize $2^{n-1}$-Dimensional Spinor
Braiding Statistics in Paired Quantum Hall States}

\author{Chetan Nayak\foot{Research supported in part by a Fannie
and John Hertz Foundation fellowship.~~~
nayak@puhep1.princeton.edu}}
\centerline{{\it Department of Physics }}
\centerline{{\it Joseph Henry Laboratories }}
\centerline{{\it Princeton University }}
\centerline{{\it Princeton, N.J. 08544 }}
\vskip .2cm
\author{Frank Wilczek\foot{Research supported in part by DOE grant
DE-FG02-90ER40542.~~~wilczek@sns.ias.edu}}
\vskip.2cm
\centerline{{\it School of Natural Sciences}}
\centerline{{\it Institute for Advanced Study}}
\centerline{{\it Olden Lane}}
\centerline{{\it Princeton, N.J. 08540}}
 
\endpage
 
\abstract{By explicitly identifying a basis valid for any number of 
electrons, we demonstrate that simple
multi-quasihole wavefunctions for the $\nu=1/2$ Pfaffian paired Hall state
exhibit an exponential degeneracy at fixed positions.  Indeed, we 
conjecture that for $2n$ quasiholes the
states realize a spinor representation of an
expanded (continuous) nonabelian statistics group $SO(2n)$.
In the four quasihole case, this is supported
by an explicit calculation of the corresponding
conformal blocks in the $c={1\over2}+1$
conformal field theory.
We present an
argument for the universality of this result,
which is significant for
the foundations of 
fractional statistics generally.  We note, for annular geometry, 
an amusing analogue to black
hole entropy.  We predict, as a generic consequence, glassy behavior.
Many of our considerations also apply to a form of the (3,3,1) state.}
 
\endpage
 
\REF\mooreread{G. Moore and N. Read, Nucl. Phys. {\bf B360} (1991)
362.
See also M. Milovanovic and N. Read, {\it Edge Excitations of Paired
Fractional Quantum Hall States}, Yale preprint (1995).}

\REF\blokwen{B. Blok and X.-G. Wen, Nucl. Phys. {\bf B374} (1992) 615.}

\REF\ho{T.L. Ho, Phys. Rev. Lett. {\bf 75} (1995) 1186.}

\REF\gww{M. Greiter, X.-G. Wen, and F. Wilczek,
Nucl. Phys. {\bf B374} (1992) 567.}

\REF\arswz{D. Arovas, J. R. Schrieffer, F. Wilczek, and A. Zee,
Nucl. Phys. {\bf B251} [FS13] (1985) 117.}

\REF\witten{E. Witten, Comm. Math. Phys. {\bf 121} (1989) 351.}

\REF\asw{D. Arovas, J.R. Schrieffer, F. Wilczek,
Phys. Rev. Lett. {\bf 53} (1984) 722.}

\REF\itzub{C. Itzykson and J.-B. Zuber, Phys. Rev. {\bf D15}
(1977) 2875.}

\REF\bpz{A.A. Belavin, A.M. Polyakov, and A.B. Zamolodchikov,
Nucl. Phys. {\bf B241} (1984) 333.}

\REF\mooreseiberg{G. Moore and N. Seiberg, Phys. Lett.
{\bf B220} (199) 422.}

\REF\coset{P. Goddard, A. Kent, and D. Olive, Comm. Math. Phys.
{\bf 103} (1986) 105.}

\REF\wzspinors{For a physically oriented account see
for example F. Wilczek and A. Zee,
Phys. Rev. {\bf D25}, 553 (1982), especially the Appendix.}

\REF\bala{A. Balachandran, L. Chandar, and A. Momen, hep-th/9512047.}

\REF\flt{M.P.A. Fisher, A.W.W. Ludwig, and A.M. Tikofsky,
talk presented at the 1996 March Meeting of the American Physical
Society and personal communication.}

\REF\sondhi{S.L. Sondhi, A. Karlhede, S.A. Kivelson, and E.H. Rezayi,
Phys. Rev. {\bf B 47} (1993) 16419.}

\REF\bloom{K. Moon, H. Mori, K. Yang, S.M. Girvin,
A.H. MacDonald, L. Zheng, D. Yoshioka, Phys. Rev. {\bf B 51}
(1995) 5138.}

\REF\gross{D.J. Gross, Nucl. Phys. {\bf 132}
(1978) 439.}

\REF\multiskyr{K. Yang and S. Sondhi, {\it Many skyrmion
wavefunctions and skyrmion statistics in quantum Hall
ferromagnets}, Princeton University preprint,
cond-mat/9605054.}

\REF\ourtextures{C. Nayak and F. Wilczek,
{\it Quantum Numbers of Textured Hall Effect Quasiparticles},
PUPT 1581, IASSNS 95/104, cond-mat/9512061.}

\REF\wzj{X. G. Wen and A. Zee, Phys. Rev.  Lett. {\bf 69} (1992) 1811.}

\chapter{Introduction}

The existence of
exotic quantum statistics in the quantum Hall effect is well
established theoretically for the classic Laughlin
states and their hierarchical descendents.  
In these cases the statistics of the quasiparticles is
one-dimensional (anyons).  
It is
a fascinating question, whether more exotic possibilities for quantum
statistics might occur in nature, and what might be their physical
consequences.
In a pioneering paper Moore and Read proposed, on the basis of subtle
arguments from conformal field theory, that quasiparticles in the
incompressible $\nu=1/2$ Pfaffian paired Hall state obey
non-Abelian statistics [\mooreread]. 
Blok and Wen considered other, more complicated,
examples [\blokwen].  Here we shall extend, and
we hope also clarify, the analysis of multi-quasihole states at $\nu=1/2$.
We shall also argue that important features of the analysis apply to
a refined version of the
$(3,3,1)$ state, and to a class of interpolating states.  In elucidating 
this application, we shall address a paradox posed by Ho
[\ho].

We have been able to analyze model wave functions for
these objects explicitly, so as to bring out their surprising
mathematical structure: they not only realize a nonabelian
representation of the braid group, 
but also naturally support a continuous extension --
the spinor representation of $SO(2n)\times U(1)$ -- of this group.
We shall discuss this analysis in the next few sections, first 
building up from small numbers of quasiholes to the general result in an
elementary but rigorous fashion, then suggesting a heuristic physical
picture. In the case of the two degenerate four quasihole
states, we give strong arguments in favor of the
proposed statistics by finding the corresponding conformal blocks
which make the braiding properties manifest.

We will then argue that the most important
qualitative features extracted from the idealized wave functions are
universal. A key step involves interpreting the relevant conformal
field theory as a strong-coupling fixed point of the renormalization group.
This logic is of interest, we believe, even for the classic
hierarchical states. 

The nonabelian statistics enforces a massive
degeneracy of states, many of which differ only with respect to subtle
high order correlations.  We expect therefore slow approach to
equilibrium, {\it i.e}. glassy behavior, as a
generic qualitative consequence.    
Since large assemblages of quasiholes carve out real-space holes,
an interesting analogue of black hole entropy is associated with the
`black disc' at the center of a paired Hall annulus.

\chapter{Ground State and Generalities}

In this and the immediately following sections
we shall construct the space of states in which
there are $2n$ quasiholes at fixed positions in
the $\nu = 1/2$ Pfaffian paired Hall state, taking the
standard trial wavefunctions literally.  These trial wavefunctions are
known to be exact zero-energy states of
the three-body Hamiltonian:
$$H = {\sum_i}\,{\sum_{j\neq i}}\,{\sum_{k\neq i,j}}\,
\delta'({z_i}-{z_j})\delta'({z_i}-{z_k})
\eqn\threebham
$$
and appear to be approximate eigenstates
of certain quasi-realistic two-body Hamiltonians [\gww].
They have the great
advantage of being quite explicit and tractable.  We will discuss the
question of universality -- that is, of robustness of the structure we
shall uncover in the model framework against small changes
in the Hamiltonian -- in a later section.

%droplet wavefunction (ground state)
 
Let us recall that the relevant
droplet wave function for the ground state takes the form
$$
\Psi (z_j) ~=~ \prod_{j<k} (z_j - z_k)^2 \prod_j e^{- |z_j|^2/4 }
  \cdot {\rm Pf~}\Biggl( {1\over z_j - z_k }\Biggr)~.
\eqn\grdstate
$$
In this equation the last factor is the Pfaffian: one chooses a
specific ordering $z_1, z_2, ... $ of the electrons, chooses a
pairing, takes the product of the
indicated factor for all pairs in the chosen pairing, and
finally takes the sum over all pairings, with the overall sign determined
by the evenness or oddness of the order in which the $z$s appear.
The result is a totally antisymmetric function.  For example for four
electrons the Pfaffian takes the form
$$
{1\over z_1 -z_2} {1\over z_3-z_4}
+ {1\over z_1 - z_3 } { 1\over z_4- z_2}
+ {1\over z_1 - z_4 } {1\over z_2 - z_3}~.
$$
This state is reminiscent of the real-space form
of the BCS pairing wavefunction; it is the
quantum Hall incarnation of a $p$-wave superconducting
state. As in a superconductor, there are half-flux
quantum excitations. The state
$$
{\Psi_{2\,{\rm qh}}} ~=~ \prod_{j<k} (z_j - z_k)^2 \prod_j e^{- |z_j|^2/4 }
  \cdot {\rm Pf~}\Biggl( {  { (z_j - \eta_1) (z_k - \eta_2 ) + 
(z_j - \eta_2) (z_k - \eta_1 )} 
\over {z_j - z_k} }\Biggr)~
\eqn\twoqhstate
$$
has half-flux quantum quasiholes at $\eta_1$
and $\eta_2$. These excitations have charge $e/4$.

One includes $2n$ quasiholes at points $\eta_{\alpha}$ 
by modifying the
Pfaffian in the manner
$$
{\rm Pf~ }\Biggl( {1\over z_j - z_k }\Biggr) \rightarrow
{\rm  Pf~  }\Biggl( { (z_j - \eta_\alpha) (z_j - \eta_\beta ) ...(z_k
-\eta_\rho ) (z_k - \eta_\sigma ) ... + (j \leftrightarrow k )
\over z_j - z_k}\Biggr)~.
\eqn\qhwf
$$
In understanding this expression it is necessary to realize that the 
$2n$ quasiholes have been divided into two groups of $n$ each
({\it i. e}. here $\alpha, \beta, ... $ and $\rho, \sigma, ... $), such
that the quasiholes within each group always act on the same electron
coordinates within an electron pair.  There are apparently
$$
{(2n)! \over {2\, n! n!} }
$$
ways of making such a division; the factor $1/2$ arising from the
possibility to 
swap the two groups of $n$ as wholes. 
Our immediate goal is to demonstrate that after linearly dependencies
are taken into account the true dimension of this space of wave
functions is actually $2^{n-1}$, and to exhibit a simple canonical basis.

\chapter{Four Quasiholes}

%elementary identity (2 electrons)

Consider first the case of four quasiholes. 
The basic identity that has to be taken into account is, in its most
primitive form,
$$
\eqalign{
(z_1 - \eta_1 )(z_1 - \eta_2 )(z_2 - \eta_3 ) (z_2 - \eta_4 ) & -
(z_1 - \eta_1 )(z_1 - \eta_3 )(z_2 - \eta_4 ) (z_2 - \eta_2 ) + 
(z_1 \leftrightarrow z_2) \cr 
&= (z_1-z_2)^2 (\eta_1 - \eta_4)(\eta_2 -\eta_3)~.\cr }
\eqn\basicid
$$
It will be convenient to abbreviate the left-hand side to 
$(12)(34) - (13)(24)$.   Then we have as an immediate consequence of
\basicid\  the relation
$$
{(12)(34) - (13)(24) \over (12)(34) - (14)(23) } ~=~ 
{(\eta_1 - \eta_4 )(\eta_2 - \eta_3 ) \over (\eta_1 - \eta_3) 
(\eta_2 - \eta_4 )} ~.
\eqn\xratio
$$
It is interesting that on the right-hand side the basic projective
invariant of four complex numbers, the cross-ratio, appears.   For
present purposes, however, the important point is 
simply that it is independent
of the $z$s.   An immediate consequence is that for two electrons and
four quasiholes the three apparently different ways of constructing
quasihole states are reduced to two through the relation
$$
(12)(34)(\eta_1 - \eta_2 )(\eta_3 - \eta_4) 
+ (13)(42) (\eta_1 - \eta_3)(\eta_4 - \eta_2) 
+ (14)(23) (\eta_1 - \eta_4)(\eta_2 - \eta_3) ~=~ 0~.
\eqn\linrel
$$
 
%many electrons

Now we want to argue that \xratio\ and \linrel\ still hold good 
for any even number of electrons, $N_e$.  To
see this we insert \basicid\ into the
Pfaffian of \qhwf : 
$$\eqalign{{{\rm Pf}_{(13)(24)}} &= {\cal A}\,\Biggl(
{{(13)(24)}\over{{z_1}-{z_2}}}\,{{(13)(24)}\over{{z_3}-{z_4}}}\,
\ldots\Biggr)\cr
&={\cal A}\,\Biggl(
{{ (12)(34) - (z_1 - z_2)^2 {\eta_{14}} {\eta_{23}}}
\over{{z_1}-{z_2}}}\,\,\,
\,{{ (12)(34) - (z_3 - z_4)^2 {\eta_{14}} {\eta_{23}}}
\over{{z_3}-{z_4}}}\,\,
\ldots\Biggr)\cr
}\eqn\manybasid$$
where ${\eta_{ij}}\equiv {\eta_i}-{\eta_j}$ and ${\cal A}$ denotes the
instruction to antisymmetrize on the $z$s.
If we expand,
$$\eqalign{ {\cal A}\,\Biggl(&
{{ (12)(34) - (z_1 - z_2)^2 {\eta_{14}} {\eta_{23}}}
\over{{z_1}-{z_2}}}
\,\,\,
\,{{ (12)(34) - (z_3 - z_4)^2 {\eta_{14}} {\eta_{23}}}
\over{{z_3}-{z_4}}}\,\,\,
\ldots\Biggr) \cr &=
{\cal A}\,\Biggl(
{{ (12)(34)}
\over{{z_1}-{z_2}}}
\,\,\,
\,{{ (12)(34)}
\over{{z_3}-{z_4}}}\,\,\,
\ldots\Biggr)\cr
&\qquad-{\cal A}\,\Biggl(
(z_1 - z_2) {\eta_{14}} {\eta_{23}}\,\,\,
\,{{ (12)(34)}
\over{{z_3}-{z_4}}}\,\,\,
\ldots\Biggr)\cr
&\qquad+
{\cal A}\,\Biggl(
(z_1 - z_2) {\eta_{14}} {\eta_{23}}\,\times\,
(z_3 - z_4) {\eta_{14}} {\eta_{23}}\,\times\,
{{ (12)(34)}\over{{z_5}-{z_6}}}\,\,\,
\ldots\Biggr) +\, \ldots\cr
}\eqn\manybasidexp$$
there will be terms on the right hand side of
\manybasidexp\ with zero, one, two, \dots, $N_e$ factors
of $(~{z_i}~-~{z_j}~)$. Upon antisymmetrization, however,
a term with $k$ factors of $(z_i - z_j)$ would have to
antisymmetrize $2k$ variables with a polynomial that
is linear in each. Since this is impossible for $k>1$,
such terms vanish. Hence
$$\eqalign{ {\cal A}\,\Biggl(&
{{ (12)(34) - (z_1 - z_2)^2 {\eta_{14}} {\eta_{23}}}
\over{{z_1}-{z_2}}}
\,\,\,
\,{{ (12)(34) - (z_3 - z_4)^2 {\eta_{14}} {\eta_{23}}}
\over{{z_3}-{z_4}}}\,\,\,
\ldots\Biggr) \cr &=
{\cal A}\,\Biggl(
{{ (12)(34)}
\over{{z_1}-{z_2}}}
\,\,\,
\,{{ (12)(34)}
\over{{z_3}-{z_4}}}\,\,\,
\ldots\Biggr)\cr
&\qquad-{\cal A}\,\Biggl(
(z_1 - z_2) {\eta_{14}} {\eta_{23}}\,\,\,
\,{{ (12)(34)}
\over{{z_3}-{z_4}}}\,\,\,
\ldots\Biggr)~.\cr}\eqn\manybasidexps$$
Similarly, one has
$$\eqalign{{{\rm Pf}_{(14)(23)}} &=
{\cal A}\,\Biggl(
{{ (12)(34)}
\over{{z_1}-{z_2}}}
\,\,\,
\,{{ (12)(34)}
\over{{z_3}-{z_4}}}\,\,\,
\ldots\Biggr)\cr
&\qquad+{\cal A}\,\Biggl(
(z_1 - z_2) {\eta_{13}} {\eta_{24}}\,\,\,
\,{{ (12)(34)}
\over{{z_3}-{z_4}}}\,\,\,
\ldots\Biggr)~.\cr}\eqn\manybasidexpss$$ From 
these we deduce the many-electron generalization
of \basicid:
$${{\rm Pf}_{(12)(34)}}-{{\rm Pf}_{(14)(23)}}=
{ {{\eta_{14}}{\eta_{23}}}\over{{\eta_{13}}{\eta_{24}}} }
\,\,\biggl({{\rm Pf}_{(12)(34)}}-{{\rm Pf}_{(13)(24)}}
\biggr)~.\eqn\fourqhiden$$
This is a linear relation among the three pairing 
possibilities for two quasiholes.
It depends on their coordinates but -- remarkably --
takes the same
form for any number of electrons.

\chapter{Six Quasiholes}

For six quasiholes, we find
the identity:
$$\eqalign{&
-\,\,{ {{\eta_{13}}{\eta_{46}}}\over{{\eta_{16}}{\eta_{34}}} }
\,\,\biggl({{\rm Pf}_{(124)(356)}}-{{\rm Pf}_{(135)(246)}}
\biggr)\,\,-\,\,
{ {{\eta_{23}}{\eta_{45}}}\over{{\eta_{25}}{\eta_{34}}} }
\,\,\biggl({{\rm Pf}_{(124)(356)}}-{{\rm Pf}_{(145)(236)}}
\biggr)\,\,+\,\,\cr &
{ {{\eta_{23}}{\eta_{46}}}\over{{\eta_{26}}{\eta_{34}}} }
\,\,\biggl({{\rm Pf}_{(124)(356)}}-{{\rm Pf}_{(146)(235)}}
\biggr)\,\,+\,\,
{ {{\eta_{13}}{\eta_{45}}}\over{{\eta_{15}}{\eta_{34}}} }
\,\,\biggl({{\rm Pf}_{(124)(356)}}-{{\rm Pf}_{(136)(245)}}
\biggr) = 0\cr}
\eqn\sixqhiden$$
in an obvious notation.  That there should be an identity of this
general type follows by arguments similar to those in the previous section,
which will be generalized in the next section.  The specific form of
the coefficients was identified using {\it Mathematica}.

Using this identity, and others related to it
by permutations, one can express all ten states in terms of just four:
${\rm Pf}_{(135)(246)}, 
{\rm Pf}_{(136)(245)}, {\rm Pf}_{(145)(236)}, {\rm Pf}_{(146)(235)}$. 
These are the four in which $\eta_{2k-1}$ never appears in the same
macrogrouping with
$\eta_{2k}$.

One can check that the remaining four states are independent.
The determinant of the three linear equations whose
solution yields \sixqhiden\ must vanish if these
states are not independent. This determinant is:
$$-2({\eta_1} - {\eta_2}) ({\eta_2 }-{ \eta_3}) {({\eta_3 }-{ \eta_4})^2} 
({\eta_1 }-{ \eta_5}) ({\eta_2} - {\eta_5}) ({\eta_1 }-{ \eta_6}) 
({\eta_4 }-{ \eta_6}) ({\eta_5 }-{ \eta_6})$$
The asymmetry is due to the arbitrary choice of
which of four variables is scaled out of the
set of three homogeneous equations.

\chapter{2n Quasiholes}

\section{Definition of the Preferred Basis}

To facilitate the analysis of the $2n$ quasihole
case, it will be convenient to have a canonical
set of $2^{n-1}$ states that -- as we will
demonstrate below -- form a basis on the space of
$2n$ quasihole states. The different
quasihole wavefunctions of the type \qhwf\ can be written
$(\eta_\alpha \eta_\beta ... )(\eta_\rho \eta_\sigma ...) $ in an
obvious notation.  Let us call these two groupings the macro-groupings.  
We can choose once and for all a reference
pairing of the $2n$ quasiholes into $n$ groups of 2.

We take as our candidate, preferred basis the set of
all macro-groupings such that the
two members of a reference
pair never belong to the same half of a macro-grouping.
To be even more specific, we will choose the reference pairs to be 
the $\eta_{2j-1}$ and $\eta_{2j}$ for $j= 1, ... , n$.  

After this we are left with $2^{n-1}$ basis states --
there are $2^n$ allowed macro-groupings, but those that differ by
interchange of the halves do not represent distinct states.
Let us call these states $\Psi_a$, $a=1,2,\ldots,{2^{n-1}}$,
$$\Psi_a = {\rm Pf}\biggl({{S_a}({z_i},{z_j})\over{{z_i}-{z_j}}}
\biggr)\eqn\sadef$$
where $S_a$ is the symmetric polynomial in ${z_i}$, ${z_j}$
corresponding to the macro-grouping $a$.
Consider one of the ${(2n)!/( 2 n! n!) } - {2^{n-1}}$
macro-groupings, $\chi$, which is not in this preferred
set. Suppose the symmetric polynomial corresponding to
$\chi$ is ${S_\chi}$.
Take the ${2^{n-1}}$ differences ${S_\chi}-{S_a}$.
Each of these differences will be of the form:
$${S_\chi}({z_1},{z_2})-{S_a}({z_1},{z_2})
= ({z_1}-{z_2})\,{P^a_{n-1}}({z_1},{z_2})
\eqn\chiadiff$$
where ${P^a_{n-1}}$ is an antisymmetric polynomial
of order $n-1$ in ${z_1}$ and ${z_2}$. To see
this, observe that the left-hand-side of
\chiadiff\ must be symmetric in ${z_1}$, ${z_2}$
since ${S_\chi}$ and ${S_a}$ are both symmetric.
Combining this with the fact that 
${S_\chi}({z_1},{z_1})={S_a}({z_1},{z_1})$,
we see that ${P^a_{n-1}}$ has the asserted
properties. If we substitute this in the expression
for $\Psi_a$, we find:
$$\eqalign{{\Psi_a} = &{\cal A}\,\Biggl(
{{S_a}\over{{z_1}-{z_2}}}\,{{S_a}\over{{z_3}-{z_4}}}\,
\ldots\Biggr)\cr
=&{\cal A}\,\Biggl(
{{{S_\chi}+({z_1}-{z_2})\,{P^a_{n-1}}}
\over{{z_1}-{z_2}}}\,\,\,
\,{{ {S_\chi}+({z_3}-{z_4})\,{P^a_{n-1}} }
\over{{z_3}-{z_4}}}\,\,
\ldots\Biggr)\cr
=&{\cal A}\,\Biggl(
{{{S_\chi}}
\over{{z_1}-{z_2}}}
\,\,\,
\,{{{S_\chi} }
\over{{z_3}-{z_4}}}\,\,\,
\ldots\Biggr)\cr
\qquad+&{\cal A}\,\Biggl(
{P^a_{n-1}}({z_1},{z_2}) \,\,\,
\,{{{S_\chi} }
\over{{z_3}-{z_4}}}\,\,\,
\ldots\Biggr)\cr
\qquad+
&{\cal A}\,\Biggl(
{P^a_{n-1}}({z_1},{z_2})\,\times\,
{P^a_{n-1}}({z_3},{z_4})\,\times\,
{{{S_\chi}}\over{{z_5}-{z_6}}}\,\,\,
\ldots\Biggr) +\, \ldots ~~.\cr
}\eqn\psichisub$$

Now consider, in \psichisub ,  
the terms with $k$ ${P^a_{n-1}}$'s. When
antisymmetrized, such terms will have factors
which are antisymmetric polynomials of order $n-1$
in $2k$ variables. Such a polynomial will be
non-zero only if $k<n/2$. Hence, \psichisub\
will have terms with zero, one, two, \dots, $[n/2]$
${P^a_{n-1}}$'s. The terms with $l$ ${P^a_{n-1}}$'s
will be antisymmetric polynomials of order
$n-1$ in $2l$ variables; there will be $n\choose 2l$
terms ${\cal A}({z_1^{p_1}}{z_2^{p_2}}\ldots{z_{2l}^{p_2l}})$
in the expansion of such a polynomial.

Now consider the equation:
$$\sum\,{c_a}\Bigr({\Psi_a}-\chi\Bigl) = 0.\eqn\psisum$$
This actually represents
$${\sum_{l=1}^{[n/2]}} {n\choose 2l} = {2^{n-1}} -1$$
equations since there are precisely this many
terms in the expansions of the surviving polynomials
in \psichisub. Since there are also ${2^{n-1}}$
${c_a}$'s, non-zero ${c_a}$'s may be found for generic
$\eta$'s such that \psisum\ holds. 

If $\sum_a c_a \neq 0$, we will have expressed $\chi$ in terms
of our putative basis states.  On the other hand, if $\sum_a c_a =0$
for non-trivial $c_a$, then \psisum\ expresses a dependence among our
putative basis states.  Thus to complete the proof of completeness it
suffices to demonstrate independence; to this demonstration we now turn.

\section{Independence of the Preferred Basis}

To prove that these states are all independent,
we must show that the equation
$$ {\sum_{a=1}^{2^{n-1}}}\,{c_a}\,{\psi_a} = 0
\eqn\dependency$$
is satisfied only if ${c_a}=0$ for all $a$.
To show this, we will exploit our freedom to move
the $z$'s in order to set all of the ${\psi_a}$'s
equal to zero except for one, ${\psi_{a^*}}$.  Then for the equation
\dependency\ to be satisfied, we must have ${c_{a*}}=0$.
Since this can be done for any ${\psi_{a^*}}$,
it will follow that the ${\psi_a}$'s
are linearly independent.

The ${c_a}$'s can be taken to be polynomials
in the $\eta_\alpha$'s, since they arise by solving linear equations in
powers of these variables.   Hence if we can show that they vanish
in some open set in ${C^{2n}}=\{({\eta_1},\ldots,{\eta_{2n}})\}$,
then we will know they vanish for arbitrary $\eta_\alpha$'s.
With this in mind, we
can focus on the special case in which the separations between 
$({\eta_1},{\eta_3})$,$({\eta_5},{\eta_7})$,
$({\eta_9},{\eta_{11}})$,..., $({\eta_{4k+1}},{\eta_{4k+3}})$,
etc. are much smaller than their distances  to any other quasiholes.
We have, for ease of presentation, supposed that the number of
quasiholes is a multiple of four.

If the members of a pair, say $\eta_1$ and $\eta_3$,
occur in opposite  macro-groupings for the
state $\psi_a$, then $\psi_a$,
considered as a function of any $z_i$, will have a zero
close to the average of $\eta_1$ and $\eta_3$.
In the limit in which we can consider all  the
other $z_j$'s to be far away, this zero will be at
${z_i} = ({\eta_1}+{\eta_3})/2$. 
If $\eta_1$ and $\eta_3$
are on the same side of the macro-grouping, however, 
then there need not be a zero near $\eta_1$ and $\eta_3$.

Let us assume that there are at least half as many electrons as
quasiholes.
Then, we can put some  $z_i$'s near the zeroes of all of the
states $\psi_a$ except ${\psi_{a^*}} = (1357\ldots)(2468\ldots)$:
specifically, we can put $z_k$ appropriately near $\eta_{4k-3}$ and
$\eta_{4k-1}$.  
Since the zeroes of the $\psi_a$'s lie near 
$\eta_1$ and $\eta_3$; $\eta_5$ and $\eta_7$;
$\eta_9$ and $\eta_{11}$;...; $\eta_{4k+1}$ and $\eta_{4k+3}$;
etc. and the zeroes of ${\psi_{a^*}}$ do not, one expects -- and we
will demonstrate momentarily -- that there
is no danger of accidentally putting one of the
$z_i$'s at a zero of ${\psi_{a^*}}$. Hence, for this
configuration of $z_i$'s (and any configuration
of the $\eta_\alpha$'s such that the inter-pair
distance is much larger than the intra-pair
distance), ${\psi_{a}}$ can be made as small as we please
for all $a\neq{a^*}$.
Hence, by our earlier arguments, ${c_{a^*}}=0$.
By considering other pairings of the
$\eta_\alpha$'s, we can show that ${c_a}=0$
for all $a$.

It only remains, then, to show that $\psi_{a^*}\neq 0$ in the chosen
configuration of $z$'s.   We will show this by evaluating it in a
special case.  Let us suppose that the number of electrons is exactly
half the number of quasiholes and that all the 
$\eta_{2r}$ are very large compared to $\eta$'s with odd subscripts.  
Clearly, if we can show that the wave function does
not vanish in this case then
the general result for this or 
larger numbers of electrons will follow.  For the
present purpose we can simplify further by assuming
$\eta_{4r-1} = \eta_{4r-3}$, and then of
course $z_r \rightarrow \eta_{4r-1}$ for $1 \leq r \leq k$.  In this very 
special
configuration the factor inside the Pfaffian associated with
$\psi_{a^*} (z_r = \eta_{4r-1}; z^\prime_r )$ becomes the common
symmetric factor
$$
\prod_{r=1}^k  (\prod_{s=1}^k \eta_{2s} (z^\prime_r - \eta_{2s-1}))
$$
times a factor which must be antisymmetrized:
$$
{\cal A} {\prod_{r=1}^k} ~ {1\over {{z^\prime_r} - {\eta_{4k-1}}}}~.
$$
Here the antisymmetrization is over the $z^\prime$'s and the $\eta$'s
which appear.  But this primeval Pfaffian certainly does not vanish.

\section{Heuristic Interpretation}

It is profoundly significant that in this proof we needed to have a
number of electrons which goes to infinity as the number of quasiholes
does.  Indeed, for a fixed number of electrons the number of
independent $2n$ quasihole states (for fixed positions)
can grow at most as a polynomial in $n$, as one readily sees by
counting the available antisymmetric polynomials.  Indeed it would be quite
strange from a physical point of view, if a finite number of
particles could support an arbitrarily 
large gas of quasiparticles with a finite entropy per quasiparticle.  
In a system of fixed size, there is a finite number
of states in the lowest Landau level at any filling
fraction. If there are $N$ electrons and $2N+n$
flux tubes (ie. $2n$ quasiholes in a $\nu=1/2$
state), there are ${(2N+n)!}\over{(N+n)!N!}$
states in the lowest Landau level.
The multiquasihole degeneracy must,
of course, be smaller than this number.

A corollary is that the distinctions among the many degenerate
quasihole states generally involve very high-order correlation
functions among the electrons.

One can understand the degeneracy from a physical point of view as
follows.  Let us imagine adding the quasiholes two at a time, bringing
in new pairs from far away.  Each electron in a pair sees half the
existing quasiholes, say the set $\Lambda$ or its complement
$\bar \Lambda$.  Symmetry among the pairs guarantees that the
same $\Lambda$ and $\bar \Lambda$ are available to each electron.  Now
the two new quasiholes, say $A$ and $B$, can be added to an expanded 
$\Lambda , {\bar \Lambda}$ in either of two ways: $\Lambda + A, {\bar
\Lambda} + B$ or $\Lambda + B, {\bar \Lambda} + A$.  At the first step,
when $\Lambda$ and ${\bar \Lambda}$ are empty, these two choices are in
reality the same, but at any subsequent stage they are inequivalent.
This gives, by induction, $2^{n-1}$ ways of building up the $2n$
quasihole state.  One might wonder whether there might be additional
possibilities for re-defining the pre-existing  $\Lambda, \bar
\Lambda$ to have unequal numbers of members, 
and adding both  $A$ and $B$ to the smaller of them.  Our analysis
shows that this cannot happen; which corresponds to a form of
clustering or closure: one can add distant pairs, without having to
reconsider the local structure among the existing pairs.  As one draws
the quasiholes into more democratic configurations, with no clear
grouping into well-separated pairs, the factorized structure will be
lost but the counting of independent states cannot change.  

Another way of regarding the multiplicity
of quasihole states, at
fixed positions, is as follows.
We have seen that there is a
dichotomic choice associated with each addition of a pair of 
nearby quasiholes to a pre-existing set. 
The two choices may be regarded as an effective spin, that may point
up or down.  Now if one imagines moving around all the quasiholes, so
that new pairings are formed, the effective spin configurations must
transform among themselves.  The structure of the independent states
suggests that they fill out a
spinor representation; and we shall see that they do.

\chapter{Conformal Field Theory: Framework}

In this and the following two sections we shall connect our
considerations on quasiholes to conformal field theory.  This will
enable us to exploit some powerful mathematical techniques for the
model quasihole wavefunctions, and will
shed considerable light on the question  of universality.

There are many reasons to believe that the appropriate effective
theories to describe incompressible quantum Hall states are
Chern-Simons gauge theories.  Since these theories do not contain any
massless particles, they describe a unique ground state with a gap.
They have the appropriate symmetries (violating P and T) and are the
lowest dimension operators consistent with these symmetries; their low
dimension makes them relevant in the renormalization group sense. 
Finally and most decisively, the point-like vortex excitations of the
Chern-Simons theory carry the exotic quantum statistics, that
characterizes the long-distance interactions of the quasiparticles
[\arswz ].

The simplest such
term, which is present in the low-energy
theory of the generalized hierarchical states is:
$${\cal L} = {K_{IJ}} {a_\mu^I}{\epsilon^{\mu\nu\lambda}}
{\partial_\nu}{a_\lambda^J} + {a_\mu^I}{j^\mu_I}\eqn\hiercs$$
where ${j^\mu_I}$ is the current of quasiholes (and quasiparticles)
of type $I$ and the ${a_\mu^I}$'s are the fictitious
gauge fields of the low-energy theory.  In the standard hierarchical
states, the $a^I$ are Abelian fields.  As is well-known,
as a result of the Chern-Simons term,
processes which move the quasiholes
around and finally bring them back to their original
positions cause the quantum state describing these
quasiholes to be multiplied by phase factors which
depend on the linking numbers of the particle trajectories.
The state furnishes a one-dimensional representation
of the braid group. 

In general, however, one may consider the possibility
of non-Abelian Chern-Simons gauge
fields appearing in the low-energy action.
In this more general situation, there is typically
not a unique state, but
a multi-dimensional space of degenerate states,
of quasiholes of specified types with fixed positions.
The braiding of quasihole trajectories is represented
by -- possibly non-commuting -- matrices acting
on this space of degenerate states. As we shall
elaborate below, the states that we constructed above
describing $2n$
quasiholes in the Pfaffian state
are perhaps the simplest
example of such a situation -- namely non-Abelian
statistics -- in the context of the quantum Hall
effect.

In a remarkable analysis of the connection between
Chern-Simons theory and knot theory,
Witten discovered, among other things, that
the states of a 2+1-dimensional Chern-Simons theory
with quasiholes at fixed positions are equal
to the conformal blocks of correlation functions
of corresponding operators in an associated
1+1-dimensional conformal field theory [\witten]. As Moore
and Read observed [\mooreread], the conformal blocks of certain
correlation functions in conformal field theories
-- being analytic functions --
can be directly interpreted as wavefunctions of electrons
in the lowest Landau level.  From the
connection between Chern-Simons theory and conformal
field theory, it is clear that a
conformal block of a correlation
function in a conformal field theory is a
representative of a universality class of
lowest Landau level wavefunctions which
have braiding properties described by the
corresponding Chern-Simons theory.

In fact, we would like to propose that 
it is precisely these braiding properties which characterize
the universality class.   One cannot expect the conformal blocks
to describe precisely the low-lying eigenstates of any realistic
Hamiltonian that supports the corresponding quantum Hall state, 
if for no other reason then because this description could
not survive small perturbations.   
One expects that the realistic wave function will differ drastically
from the ideal one whenever two particles approach one another,
reflecting their non-universal short-range interactions.  
In conformal field theory
the wave functions have no preferred length scale. The theory has in effect
only ultra-short range interactions, encoded in the fusion rules, and
otherwise is completely encoded in the braiding properties (which are
constrained by the fusion rules).  Small local perturbations which do
not change the universality class will change the structure of the
wavefunctions as particles approach one another -- within the range of
the interaction projected onto the lowest Landau level, and therefore
(for reasonable interactions) cut off at the magnetic length -- but 
not the braiding properties involving paths such that 
particles never approach closely.

Not all conformal field theories can produce
sensible lowest Landau level wavefunctions.
A conformal block of a correlation function
in an arbitrary
conformal field theory will usually
contain poles and may not even be a single-valued
function. Even if it is a normalizable
wavefunction enforcing Fermi statistics
on the electrons, it may not be the ground
state of a positive-definite Hamiltonian.
Furthermore, it is possible that several
different conformal field theories yield the
same quantum Hall ground state but different
quasihole states and therefore different
representations of the braid group, only
one of which is realized in the quantum Hall effect.

Despite these potential difficulties, there are examples of conformal
field theories for which the correspondence with
a quantum Hall ground state and its concomitant
quasihole states (in other words, the zero-energy
states of some model Hamiltonian) is complete.
The simplest non-trivial 
example is provided by  the $c=1$ theory of a
chiral boson, $\phi$, with compactification radius
$r=\sqrt{m}$.   This yields the Laughlin state [\mooreread]:
$${\Psi_{1/m}} = \langle\, {e^{i\sqrt{m}\phi({z_1})}}\,
{e^{i\sqrt{m}\phi({z_2})}}\,\ldots\,{e^{i\sqrt{m}\phi({z_N})}}
\,{e^{-i \,\int{d^2}z\,\sqrt{m}{\rho_0}\phi(z)}}
\rangle = {\prod_{i>j}}\, {({z_i} - {z_j})^m}~,
\eqn\laughcorrfcn$$
which is known to be exact for appropriate ultra-local Hamiltonians.

The last factor in the correlation function 
corresponds to a neutralizing background ($\rho_0$ is the electron density);
without it, 
this correlation function would vanish.  Its inclusion might appear to be a
technical subtlety, but it is profoundly important.
If we
regard this factor as part of the action,
the resulting theory will be nonunitary (with $c<0$),
since the neutralizing background provides 
an imaginary term in the action.
As one consequence, the ${e^{i\sqrt{m}\phi}}$ operator
will have negative dimension; indeed 
it must, since the correlation function
increases with separation. These points
will be crucial in justifying the connection  
of conformal field theory braiding factors to 
Berry phases in the microscopic theory, formalizing the preceding
heuristic discussion.

The state with one quasihole is given by:
$$\eqalign{{\Psi_{1/m}^{\rm qh}} &= \langle\,{e^{{i\over\sqrt{m}}\phi(\eta)}}
{e^{i\sqrt{m}\phi({z_1})}}\,
{e^{i\sqrt{m}\phi({z_2})}}\,\ldots\,{e^{i\sqrt{m}\phi({z_N})}}
\,{e^{-i \,\int{d^2}z\,\sqrt{m}{\rho_0}\phi(z)}}
\rangle\cr &= {\prod_k}\,({z_k}-\eta)\,\,
{\prod_{i>j}}\, {({z_i} - {z_j})^m}~.\cr}\eqn\lqhcorrfcn$$
In other words, electrons are represented by
the operator ${e^{i\sqrt{m}\phi}}$ while
quasiholes are represented by ${e^{{1\over\sqrt{m}}\phi}}$.
Multi-quasihole states can be obtained
by a straightforward generalization of \lqhcorrfcn.
The braiding properties of electrons
may be seen by inspection from the wavefunctions \laughcorrfcn\
and \lqhcorrfcn;
the braiding properties of the
quasiholes may be obtained from a Berry phase
calculation [\asw]. These braiding properties are
precisely those described by the abelian
Chern-Simons theory associated
with the $c=1$ conformal field theory:
$${\cal L} = m {a_\mu}{\epsilon^{\mu\nu\lambda}}
{\partial_\nu}{a_\lambda} + {a_\mu}{j^\mu}~,\eqn\laughcs$$
where $j^\mu$ is the quasihole current and
an electron is simply an aggregate of
$m$ quasiparticles.

\chapter{Conformal Field Theory: Application to Pfaffian State}

We will adopt the point of view that finding the
appropriate conformal field theory allows us
to identify the Chern-Simons
theory describing the universal low-energy
properties of the quantum Hall state.  For the 
Laughlin and hierarchical states this back door approach 
is unnecessary,
since the correct Chern-Simons theory can be
found more directly -- but for the Pfaffian state it
will prove to be quite fruitful.

As Moore and Read
found [\mooreread], the Pfaffian state is
given by a correlation function in a theory
of a Majorana fermion, $\psi$, and a chiral boson, $\phi$
with $c={1\over 2} + 1$.
$$\eqalign{ {\Psi_{\rm Pf}} &= \langle \psi({z_1}),\psi({z_2})\,\ldots\,
\psi({z_N})\,\,
{e^{i\sqrt{2}\phi({z_1})}}\,
{e^{i\sqrt{2}\phi({z_2})}}\,\ldots\,{e^{i\sqrt{2}\phi({z_N})}}
{e^{-i \,\int{d^2}z\,\sqrt{m}{\rho_0}\phi({z})}}
\rangle\cr \qquad &= {\rm Pf}\biggl({1\over{{z_i}-{z_j}}}\biggr)\,\,
{\prod_{i>j}} {({z_i} - {z_j})^m}~.\cr}\eqn\pfcorrfcn$$
Evidently, the electron is represented by the
operator $\psi{e^{i\sqrt{2}\phi}}$.
They also found the two-quasihole state:
$$\eqalign{ {\Psi_{\rm Pf}^{\rm 2 qh}} &=
\langle \sigma({\eta_1})\,\sigma({\eta_2})\,
\psi({z_1}),\psi({z_2})\,\ldots\,
\psi({z_N})\,\,\times\cr \qquad\qquad\qquad\qquad
&\,{ e^{{i\over{2 \sqrt{2}}}\phi({\eta_1})} }
{e^{{i\over{2 \sqrt{2}}}\phi({\eta_2})}}
{e^{i\sqrt{2}\phi({z_1})}}\,
{e^{i\sqrt{2}\phi({z_2})}}\,\ldots\,{e^{i\sqrt{2}\phi({z_N})}}
{e^{-i \,\int{d^2}z\,\sqrt{m}{\rho_0}\phi({z})}}
\rangle\cr &= {\rm Pf}\biggl({1\over{{z_i}-{z_j}}}\, 
\Bigl( {({z_i}-{\eta_1})({z_j}-{\eta_2}) + (i\leftrightarrow j)}
\Bigr) \biggr)\,\,
{\prod_{i>j}} {({z_i} - {z_j})^m} \cr}\eqn\pftwoqhcorr$$
where $\sigma$ is the twist field of the fermion or,
equivalently, the spin field in the Ising model interpretation
of the $c={1\over 2}$ theory.
The choice of the operator
$\sigma {e^{{i\over{2 \sqrt{2}}}\phi}}$ to
represent the quasihole is dictated by the condition
that the wavefunction \pftwoqhcorr\ be single-valued
in the electron coordinates.
They were unable to find an explicit form for the four-quasihole
states, but observed that the conformal
blocks of the corresponding correlation function
$$\eqalign{ {\Psi_{\rm Pf}^{\rm 4 qh}} &=
\langle \sigma({\eta_1})\,\sigma({\eta_2})\,
\sigma({\eta_3})\,\sigma({\eta_4})\,
\psi({z_1}),\psi({z_2})\,\ldots\,
\psi({z_N})\,\,\times\cr 
&\qquad
{e^{{i\over{2 \sqrt{2}}}\phi({\eta_1})}}
{e^{{i\over{2 \sqrt{2}}}\phi({\eta_2})}}
{e^{{i\over{2 \sqrt{2}}}\phi({\eta_3})}}
{e^{{i\over{2 \sqrt{2}}}\phi({\eta_4})}}\,\times\cr
&\qquad {e^{i\sqrt{2}\phi({z_1})}}\,
{e^{i\sqrt{2}\phi({z_2})}}\,\ldots\,{e^{i\sqrt{2}\phi({z_N})}}
{e^{-i \,\int{d^2}z\,\sqrt{m}{\rho_0}\phi({z})}}
\rangle\cr}\eqn\pffourqhcorr$$
form a two-dimensional vector space. This is
the same degeneracy which we found in the previous section.

To make a more direct comparison with the
wavefunctions of the previous section, 
we will need the explicit
form of these conformal blocks.
To calculate them, we use the bosonization
approach of Itzykson and Zuber [\itzub],
which takes advantage of the fact that two independent
copies of the $c={1\over 2}$
theory are equivalent to a $c=1$ boson.
If we call the spin and Majorana fermion fields
of our two $c={1\over 2}$ theories $\sigma_1$,
$\sigma_2$ and $\psi_1$, $\psi_2$, then
we have the following bosonization formulas:
$${\sigma_1}{\sigma_2} \rightarrow \cos{\varphi\over 2}
\eqn\sigmabose$$
$${\psi^{L,R}_1}
\rightarrow \cos{\varphi_{L,R}}\eqn\psiobose$$
$${\psi^{L,R}_2}
\rightarrow \sin{\varphi_{L,R}}~. \eqn\psitbose$$
The four-quasihole correlation function can be obtained
from:
$$\langle {\sigma_1}\,{\sigma_1}\,{\sigma_1}\,{\sigma_1}\,
{\psi^R_1}{\psi^R_1}\ldots{\psi^R_1}\rangle\,
\langle {\sigma_2}{\sigma_2}{\sigma_2}{\sigma_2}\rangle =
\langle \cos{\varphi\over 2}\,\cos{\varphi\over 2}\,
\cos{\varphi\over 2}\,\cos{\varphi\over 2}\,
\cos{\varphi_R}\ldots\cos{\varphi_R}\rangle
\eqn\fourqhcorr$$
and
$$\langle {\sigma_1}\,{\sigma_1}\,{\sigma_1}\,{\sigma_1}\rangle
\langle {\sigma_2}\,{\sigma_2}\,{\sigma_2}\,{\sigma_2}\rangle =
{\langle {\sigma_1}\,{\sigma_1}\,{\sigma_1}\,{\sigma_1}\rangle^2} =
\langle \cos{\varphi\over 2}\,\cos{\varphi\over 2}\,
\cos{\varphi\over 2}\,\cos{\varphi\over 2} \rangle ~.
\eqn\fqhauxcorr$$
Equations \fourqhcorr\ and \fqhauxcorr\ may
be evaluated using Wick's theorem. Combining them,
we find:
$$\eqalign{ \langle {\sigma_1}\,{\sigma_1}\,{\sigma_1}\,{\sigma_1}\,
{\psi^R_1}{\psi^R_1}\ldots{\psi^R_1}\rangle\, &= 
{1\over{|1+\sqrt{1-x}| + |1-\sqrt{1-x}|}}\,\,
{\biggl|{{{\eta_{13}}{\eta_{24}}}
\over{{\eta_{12}}{\eta_{23}}{\eta_{34}}{\eta_{41}}}}\biggr|^{1\over4}}
\,\times\cr& \qquad
\Biggl( {\rm Pf}\biggl({1\over{{z_i}-{z_j}}}\,{h_{(13)(24)}}
\biggr)\,+\,|1-x|\,\,
{\rm Pf}\biggl({1\over{{z_i}-{z_j}}}\,{h_{(14)(23)}}\biggr)
\,+\cr & \qquad\qquad\qquad |x|\,\,
{\rm Pf}\biggl({1\over{{z_i}-{z_j}}}\,{h_{(12)(34)}}\biggr)\Biggr)\cr}
\eqn\fourqhexp$$
where we have abbreviated
${\eta_{\alpha\beta}}={\eta_\alpha}-{\eta_\beta}$,
$x$ is the cross-ratio, or anharmonic ratio:
$$x ={ {({\eta_1}-{\eta_2})({\eta_3}-{\eta_4})}\over
{({\eta_1}-{\eta_3})({\eta_2}-{\eta_4}) }}$$
and
$${h_{(\alpha\beta)(\gamma\delta)}} = 
{\Biggl({{({z_i}-{\eta_\alpha})({z_i}-{\eta_\beta})}
\over{({z_i}-{\eta_\gamma})({z_i}-{\eta_\delta})}}\,\,
{{({z_j}-{\eta_\gamma})({z_j}-{\eta_\delta})}
\over{({z_j}-{\eta_\alpha})({z_j}-{\eta_\beta})}}\Biggr)^{1\over2}}
\,\,+ \,\, (i\leftrightarrow j) ~.\eqn\hdef$$
We are interested in the conformal blocks,
${{\cal F}_0^L}$, ${{\cal F}_0^R}$, ${{\cal F}_{1/2}^L}$,
${{\cal F}_{1/2}^L}$ of this correlation function,
$$\langle {\sigma_1}\,{\sigma_1}\,{\sigma_1}\,{\sigma_1}\,
{\psi^R_1}{\psi^R_1}\ldots{\psi^R_1}\rangle\, =
{{\cal F}_0^R}({z_i},{\eta_\alpha})
{{\cal F}_0^L}({{\bar\eta}_\beta}) +
{{\cal F}_0^R}({z_i},{\eta_\alpha})
{{\cal F}_0^L}({{\bar\eta}_\beta})\eqn\cblockdef$$
which satisfy the differential equations:
$$\Biggl({4\over3}\,{{\partial^2}\over{\partial{\eta_\alpha^2}}}
\,-\,{\sum_{\beta\neq\alpha}}\,\biggl(
{{1/16}\over{({\eta_\alpha}-{\eta_\beta})^2}}\,+\,
{1\over{({\eta_\alpha}-{\eta_\beta})}}\,
{\partial\over{\partial{\eta_\beta}}}\biggr)
\,-\,{\sum_{i}}\,\biggl(
{{1/2}\over{({\eta_\alpha}-{z_i})^2}}\,+\,
{1\over{({\eta_\alpha}-{z_i})}}\,
{\partial\over{\partial{z_i}}}\biggr)\Biggr)\,\,
{{\cal F}_p^R} = 0\eqn\rdiffeq$$
$$\Biggl({4\over3}\,{{\partial^2}\over{\partial{{\bar\eta}_\alpha^2}}}
\,-\,{\sum_{\beta\neq\alpha}}\,\biggl(
{{1/16}\over{({{\bar\eta}_\alpha}-{{\bar\eta}_\beta})^2}}\,+\,
{1\over{({{\bar\eta}_\alpha}-{{\bar\eta}_\beta})}}\,
{\partial\over{\partial{{\bar\eta}_\beta}}}\biggr)
\Biggr)\,\,{{\cal F}_p^L} = 0\eqn\ldiffeq$$
that follow from the degeneracy of the
the spin fields at level 2 [\bpz].

This problem simplifies considerably because the Majorana
fermions, $\psi$, are purely right-handed,
as we have already indicated in
\cblockdef\ and \ldiffeq. Hence,
these fields do not participate in the
left-handed, or anti-holomorphic,
part of the correlation function. 
The anti-holomorphic conformal blocks are
functions only of the ${\bar\eta}_\alpha$'s.
By conformal invariance, their only non-trivial
dependence is on the cross-ratio, ${\bar x}$,
so \ldiffeq\ may be reduced to an ordinary
differential equation with solutions [\bpz]:
$${{\cal F}_0^L} = 
{\biggl({{{{\bar \eta}_{13}}{{\bar \eta}_{24}}}
\over{{{\bar \eta}_{12}}{{\bar \eta}_{23}}
{{\bar \eta}_{34}}{{\bar \eta}_{41}}}}\biggr)^{1\over8}}
\,\,\Bigr(1 + \sqrt{1-{\bar x}}\Bigl)^{1/2}\eqn\lidblock$$
$${{\cal F}_{1/2}^L} = 
{\biggl({{{{\bar \eta}_{13}}{{\bar \eta}_{24}}}
\over{{{\bar \eta}_{12}}{{\bar \eta}_{23}}
{{\bar \eta}_{34}}{{\bar \eta}_{41}}}}\biggr)^{1\over8}}
\,\,\Bigr(1 - \sqrt{1-{\bar x}}\Bigl)^{1/2}~.\eqn\lpsiblock$$
Substituting these anti-holomorphic conformal blocks
into the right-hand-side of \cblockdef\
and using the expression \fourqhexp\ for the left-hand-side,
we find, by equating the coefficients
of the different functions of ${\bar x}$:
$$\eqalign{ {{\cal F}_{0,1/2}^R} &=
{1\over{(1 \pm \sqrt{1-{x}})^{1/2}}}\,
{\biggl({{{\eta_{13}}{\eta_{24}}}
\over{{\eta_{12}}{\eta_{23}}{\eta_{34}}{\eta_{41}}}}\biggr)^{1\over8}}
\,\,\times\cr &\qquad\qquad
\Biggr({\rm Pf}\biggl({1\over{{z_i}-{z_j}}}\,{h_{(13)(24)}}
\biggr) \,\,\pm\,\,\sqrt{1-x}\,
{\rm Pf}\biggl({1\over{{z_i}-{z_j}}}\,{h_{(14)(23)}}
\biggr)\Biggl)~.\cr }\eqn\fspinblocks$$
Remarkably, the consistency of
equations \fourqhexp, \cblockdef, \lidblock, and
\lpsiblock\ which allows
us to find these solutions depends
upon the identity \fourqhiden.
For small numbers of electrons,
we have checked, using
{\it Mathematica}, that \fspinblocks\ satisfy
the differential equations \rdiffeq.
Including, as well, the contribution of the $c=1$
part of the theory, we find the four-quasihole
wavefunctions:
$${\Psi^{(4\,qh, 0)}} = { {\Bigl({\eta_{13}}{\eta_{24}}\Bigr)^{1\over 4}}
\over{(1 + \sqrt{1-{x}})^{1/2}}}\,
\Biggr( {\Psi_{(13)(24)}} \,\,+\,\,\sqrt{1-x}\,\,
{\Psi_{(14)(23)}}\Biggl)
\eqn\fourqhid$$
$${\Psi^{(4\,qh, 1/2)}} = {{\Bigl({\eta_{13}}{\eta_{24}}\Bigr)^{1\over 4}}
\over{(1 - \sqrt{1-{x}})^{1/2}}}\,
\Biggr( {\Psi_{(13)(24)}} \,\,-\,\,\sqrt{1-x}\,\,
{\Psi_{(14)(23)}}\Biggl)~.
\eqn\fourqhpsi$$
These wavefunctions are linear combinations of the
ones we found earlier.

We have not been able to calculate the conformal
blocks corresponding to $2N$ quasiholes
for $N>2$. However, according to standard
arguments, there are $2^{N-1}$ conformal blocks
since each pair of spin fields can fuse
to form the identity or a fermion,
$\sigma\,\sigma \sim 1 + \psi$ and the total
number of $\psi$'s in the correlation function
must be even. We conjecture that these
conformal blocks are simply a different --
and, as we will argue below, particularly useful --
basis for the wavefunctions which we enumerated
earlier.

\chapter{Conformal Field Theory and Braiding}

To calculate the braiding properties
of the quasiholes we must the calculate Berry
phase matrix:
$${\gamma_{ij}} = i\,\int\,{dt}
\langle{\psi_i}(t)|\,{d\over{dt}}|{\psi_j}(t)\rangle
\eqn\berryphase$$
where $|{\psi_j}(t)\rangle$ is a wavefunction
in which one of the quasihole positions, ${\eta_\alpha}(t)$,
traces out a loop around
another quasihole as $t$ is varied.
These Berry phases will depend on the basis
that we choose because we can take
linear combinations of \fourqhid\ and \fourqhpsi\ with
arbitrary $\eta$-dependent coefficients.
The physical, basis independent effect of braiding
is the combination of the Berry phase matrix
and the transformation properties that are manifest
in the wavefunction. Consider the basis $\Psi_{(13)(24)}$,
$\Psi_{(14)(23)}$. If we take $\eta_1$ around $\eta_2$
in either of these wavefunctions, there is no effect.
Hence, all of the effects of braiding quasiholes is
contained in the Berry phase matrix. On the other hand,
if we take $\eta_1$ around $\eta_3$ in the
wavefunctions ${\Psi^{(4\,qh, 0)}}$,
${\Psi^{(4\,qh, 1/2)}}$, avoiding the other quasiholes,
they transform into each
other because of the branch cut in $\sqrt {\eta_1 - \eta_3}$ 
within the $\sqrt{1-x}$ factors.
Of course, the Berry phase calculation
in this basis will also 
be different than in the other basis.  But
the combined effect of the
manifest transformation and the acquired Berry phase
must be the same in both bases, since
it is physically observable in interference between amplitudes
involving such trajectories and unbraided trajectories within a path integral. 
We will now argue for the hypothesis that: {\it in
the basis specified by the conformal
blocks, there is no Berry phase and all of the
transformation properties under braiding
are manifest}.

Since the conformal blocks transform as the
states of a Chern-Simons theory, this hypothesis
must hold if the Chern-Simons theory describes
the low-energy physics of the corresponding
quantum Hall states. It was noted
in [\mooreread] and especially [\blokwen] that it 
holds in a number of cases including the Laughlin states.
Now we supply a general argument
in its favor. 

The quantities which we would
like to calculate and manipulate are integrals of the form:
$$\int \, {\prod_i} {d^2}{z_i}\,\,
{{\cal F}_p}({\eta^\prime_\alpha},{z_i})
{{\bar{\cal F}}_p}({{\bar\eta}_\alpha},{{\bar z}_i})~.
\eqn\cblockint$$
These provide effective wave functions -- more precisely, off-diagonal
density matrices -- for the quasiholes.  
For ease of notation we shall drop the primes on the $\eta'$'s;
implicitly understanding that $\eta$ and $\bar \eta$ should be regarded as
independent variables.
Such products of conformal blocks
make up correlation functions, so we can
rewrite \cblockint\ as:
$$\eqalign{\int\, {\prod_i} {d^2}{z_i}\,&\,
\langle {{\cal O}_q}({\eta_1},{{\bar\eta}_1})\ldots
{{\cal O}_q}({\eta_n},{{\bar\eta}_n})\,\,
{{\cal O}_e}({z_1},{{\bar z}_1})\ldots
{{\cal O}_e}({z_N},{{\bar z}_N})\rangle =\cr 
&{{d^N}\over{dt^N}}
\langle {{\cal O}_q}({\eta_1},{{\bar\eta}_1})\ldots
{{\cal O}_q}({\eta_n},{{\bar\eta}_n})
{e^{t\, \int\, {d^2}z\, {{\cal O}_e}(z,{\bar z}) }}
\rangle\cr}
\eqn\corrfcnint$$
where ${{\cal O}_q}$ and ${{\cal O}_e}$ are the
left-right symmetric versions of the quasihole
and electron operators.  (In the case of the $c={1\over2}+1$
theory, ${{\cal O}_q}=\sigma\,{e^{i\phi/2\sqrt{2}}}$
and ${{\cal O}_e}={\psi_L}{\psi_R}\,{e^{i\phi\sqrt{2}}}$.)

As we have discussed before, 
\corrfcnint\ as written requires that we include the
background charge operator,
${e^{-i \,\int{d^2}z\,\sqrt{m}{\rho_0}\phi(z)}}$,
in the action. Because of the background charge,
${{\cal O}_e}$ is actually an operator of
negative scaling dimension. As we remarked
earlier, this must be true if we want
the correlation function to increase with separation,
which indeed we must require for 
a lowest Landau level wavefunction.
If we now include the term 
$t\, \int\, {d^2}z\, {{\cal O}_e}(z,{\bar z})$
as a term in the action, as in \corrfcnint , this 
strongly relevant operator will drive the theory to
some strongly coupled fixed point.  However,
at a strongly coupled fixed point dominated
by potential energy terms we expect
there to be a gap, and thus we expect that 
the correlation function
\corrfcnint\ will approach a constant (possibly zero)
independent of the $\eta$'s in the limit
of well-separated $\eta$'s. 
Of course, in this circumstance
integrands appearing 
in the Berry phase calculation
\berryphase, will vanish, since the derivative acts
on an $\eta$-independent quantity. 
Therefore, in this basis the Berry phase
contribution to the transformation properties
of the wavefunctions 
are trivial.  This argument provides, in a sense,
a generalization of the plasma analogy
that is valid for any lowest Landau level
wavefunction arising as a conformal
block in a conformal field theory.

As we have seen in the preceding paragraphs,
the dimensions of the spaces of conformal blocks
in the $c={1\over2}+1$ conformal field theory
coincide with the dimensions of the
spaces of multi-quasihole states. In the case of the
four-spin conformal blocks, we have shown that these
are exactly the same 
as the four-quasihole wavefunctions (in a special basis).
It is hard to resist the conclusion that
the braiding properties of quasiholes
in the Pfaffian state are precisely
those of the Chern-Simons theory which
corresponds to the $c={1\over2}+1$ conformal field theory; and thus
that in this precise sense they embody nonabelian statistics.
 
Finally
we now display, for the record, the relevant Chern-Simons theory explicitly.
The $c=1$ part of the theory corresponds simply
to a $U(1)$ Chern-Simons field with coupling
constant ${k_{U(1)}}=2$. According to the Kac-Moody algebra
coset construction of [\coset], the $c={1\over2}$ theory is equivalent
to ${SU(2)_1}\times{SU(2)_1}/{SU(2)_2}$, where the
${SU(2)_2}$ that is modded out is the diagonal
subgroup of ${SU(2)_1}\times{SU(2)_1}$.
As Moore and Seiberg showed [\mooreseiberg], the ${G_k}/{H_{k'}}$
coset conformal field theory is equivalent
to a Chern-Simons theory with $G$ and $H$ gauge
fields with coupling constants $k$ and $-k'$,
respectively. Hence the proposed Chern-Simons theory
for the Pfaffian state and its multi-quasihole
states is:
$$\eqalign{{\cal L}\,\,\, &=\,\,\, {a^1_{\mu a}}{\epsilon^{\mu\nu\lambda}}
{\partial_\nu}{a^1_{\lambda a}}\,\,\, +\,\,\, {2\over3}\,{\epsilon^{abc}}
{\epsilon^{\mu\nu\lambda}} {a^1_{\mu a}}{a^1_{\nu b}}{a^1_{\lambda c}}
\cr & \qquad \,+\,\,
{a^2_{\mu a}}{\epsilon^{\mu\nu\lambda}}
{\partial_\nu}{a^2_{\lambda a}}\,\,\, +\,\,\, {2\over3}\,{\epsilon^{abc}}
{\epsilon^{\mu\nu\lambda}} {a^2_{\mu a}}{a^2_{\nu b}}{a^2_{\lambda c}}
\cr & \qquad \,-\,\,
2\,\biggl({a^3_{\mu a}}{\epsilon^{\mu\nu\lambda}}
{\partial_\nu}{a^3_{\lambda a}}\,\,\, +\,\,\, {2\over3}\,{\epsilon^{abc}}
{\epsilon^{\mu\nu\lambda}} {a^3_{\mu a}}{a^3_{\nu b}}{a^3_{\lambda c}}
\biggr)
\cr & \qquad \,
+\,\,  2 \,{c_\mu}{\epsilon^{\mu\nu\lambda}}
{\partial_\nu}{c_\lambda}\cr}
\eqn\pfaffcs$$
where ${a^1_{\lambda a}}$, ${a^2_{\lambda a}}$, and
${a^3_{\lambda a}}$ are $SU(2)$ gauge fields and
$c_\lambda$ is a $U(1)$ gauge field. The interaction
with quasiholes is specified by the representations of
the three $SU(2)$'s and one $U(1)$ in which
they transform. The half-flux quasiparticles
transform in the spin-$0$, ${1\over2}$, and ${1\over2}$
representations of the three $SU(2)$'s and the
charge-${1\over2}$ representation of the $U(1)$
(since it is ${1\over2}$ of a flux quantum).

\chapter{Spinor Structure}

To elucidate the $2^{n-1}$ states of $2n$ quasiholes
further, let us introduce an appropriate
notation.  The first half of the macro-grouping will contain $\eta_1$
or $\eta_2$ but not both, either $\eta_3$ or $\eta_4$ but not both,
and so forth.  For each reference pair, let us denote the appearance
of the first member (odd index) in the first half of the
macro-grouping with a + sign, and the appearance of the second member
(even index) with a - sign.  Then the final basis states are
represented by n-component arrays of the type $(\pm, \pm, ... )$, with
the understanding that arrays which differ by an overall change of the
sign of every component represent the same state.  Clearly then there
are $2^{n-1}$ states appearing on a very symmetrical basis.  Indeed,
the dimension and form of the final basis set is extremely suggestive 
of a spinor representation of $SO(2n)$, written in an $SO(2)^n$ basis 
[\wzspinors ].

We can interpret this as an expansion of the statistics group.
Indeed, the interchange of two quasiholes can be considered 
as a $\pi$ rotation {\it in configuration space} in the plane
containing their coordinates, leaving the other quasihole positions
fixed.  As we will demonstrate below, this operation lifts to an
operation on the quasihole states, in such a way that there is a
natural interpolation to a continuous action of $SO(2n)\times U(1)$,
under which our independent states transform as spinors.  

Actually the basis that we chose above, though
suggestive, is not the most convenient at this point. 
As we argued in the previous
section, the most convenient basis for
a discussion of statistics is the one given
by the conformal blocks of the
$c={1\over 2}+1$ theory. Let us define
basis states $|{\epsilon_1},{\epsilon_2},\ldots,{\epsilon_n}\rangle$
with ${\epsilon_i}=\pm 1$ and ${\prod_i}{\epsilon_i}=1$
such that $|{\epsilon_1},{\epsilon_2},\ldots,{\epsilon_n}\rangle$
is the conformal block arising from the channel
in which $\sigma({\eta_{2i-1}})\sigma({\eta_{2i}})\sim 1$
if ${\epsilon_i}=+1$ and
$\sigma({\eta_{2i-1}})\sigma({\eta_{2i}})\sim \psi$
if ${\epsilon_i}=-1$.
In this notation, ${\Psi^{(4\,qh, 0)}}$ is $|++\rangle$
and ${\Psi^{(4\,qh, 1/2)}}$ is $|--\rangle$.
As we will discuss shortly, this notation implies
that the $2n$ quasihole state space is the
representation space for the spinor representation
of $SO(2n)$.

{\it The action of an interchange
on the $2n$-quasihole state space is given
by the representatives in the spinorial
representation of $SO(2n)\times U(1)$ of
$\pi/2$ rotations in the same plane
in which the interchange is implemented by
a $\pi$ rotation in the quasiholes'
configuration space, $C^{2n}$}. This
is sufficient to specify the action of
the full braid group since it can be
generated by elementary interchanges.
The $2n$-quasihole states are a representation
space for the full $SO(2n)\times U(1)$, not merely
the subgroup which represents the braid group.
As a result, $SO(2n)\times U(1)$ can be
thought of as a continuous extension of the
braid group acting on these states.

We will verify this assertion in the four-quasihole
case with our explicit wavefunctions \fourqhid\ and
\fourqhpsi, and give an argument in favor of its
validity in the $2n$-quasihole case.

Recall that $S0(2n)$ representations are constructed
from ${\gamma_1}, {\gamma_2}, \ldots, {\gamma_{2n}}$:
$$\eqalign{
{\gamma_1} &= {\tau_1}\otimes{\tau_3}\otimes\ldots\otimes{\tau_3}\cr
{\gamma_2} &= {\tau_2}\otimes{\tau_3\otimes}\ldots\otimes{\tau_3}\cr
&\vdots\cr
{\gamma_{2k-1}} &= 1\otimes\ldots\otimes 1
\otimes{\tau_1}\otimes{\tau_3}\ldots\otimes{\tau_3}\cr
{\gamma_{2k}} &= 1\otimes\ldots\otimes 1
\otimes{\tau_2}\otimes{\tau_3\otimes}\ldots\otimes{\tau_3}\cr
&\vdots\cr
{\gamma_{2n-1}} &= 1\otimes\ldots\otimes 1
\otimes{\tau_1}\cr
{\gamma_{2n}} &= 1\otimes\ldots\otimes 1
\otimes{\tau_2}\cr
}
\eqn\gammadefs$$
where the $\tau_i$'s are the Pauli matrices.
${\gamma_{2k-1}}$ and ${\gamma_{2k}}$ have $k-1$
$1$'s to the left and $n-k$ ${\tau_3}$'s to the
right of ${\tau_1}$ and ${\tau_2}$, respectively.
The generator of rotations in the $i-j$ plane
is ${\sigma_{ij}}={1\over 4}\,i\,[{\gamma_i},{\gamma_j}]$.
The representation space is the vector space
spanned by the states
$|{\epsilon_1},{\epsilon_2},\ldots,{\epsilon_n}\rangle$
with ${\epsilon_i}=\pm 1$ and ${\prod_i}{\epsilon_i}=1$.
The $q^{\rm th}$ factor in the direct products
\gammadefs\ which make up the ${\sigma_{ij}}$'s
acts on ${\epsilon_k}=\pm 1$ as if these were
spin up and spin down. The $U(1)$ action will simply
be multiplication by $e^{i\theta/4}$. For an
interchange, this will be a factor by $e^{i\pi/4}$,
as we would expect for charge $e/4$ excitations.

The spinorial representation of $SO(4)$
is two-dimensional, let us call these
states $|++\rangle$ and $|--\rangle$.
We will verify that $|++\rangle$ and $|--\rangle$
transform in precisely the same way under
the rotation $e^{i{\pi\over 2}{\sigma_{13}}}\,{e^{i\pi/4}}$
as ${\Psi^{(4\,qh, 0)}}$ and
${\Psi^{(4\,qh, 1/2)}}$ transform under
an exchange of $\eta_1$ and $\eta_3$.  From \gammadefs\ 
and the definition of
${\sigma_{13}}$,
$$
{e^{i\pi/4}}\,
{e^{i{\pi\over 4}{\sigma_{13}}}}|++\rangle =
{{e^{i\pi/4}}\over\sqrt{2}}\bigl(|++\rangle + |--\rangle\bigr)~.
\eqn\plustransf
$$

Consider, now, an exchange of $\eta_1$ and $\eta_3$ in
the wavefunction ${\Psi^{(4\,qh, 0)}}$. Under
this exchange, $1-x\rightarrow x$, so:
$$
{\Psi^{(4\,qh, 0)}} \rightarrow {e^{i\pi/4}}\,
{{\Bigl({\eta_{13}}{\eta_{24}}\Bigr)^{1\over 4}}
\over{(1+\sqrt{x})^{1/2}}}\,\,
\Biggl({\Psi_{(13)(24)}}\,\,+\,\,\sqrt{x}\,
{\Psi_{(12)(34)}}\Biggl)~.
\eqn\psiidtransone
$$
At this point the hero of our story,
the identity \fourqhiden, enters the fray once more.
This identity expresses 
${\Psi_{(12)(34)}}$ in terms of
${\Psi_{(13)(24)}}$ and ${\Psi_{(14)(23)}}$.
Using it, we can rewrite \psiidtransone:
$$
{\Psi^{(4\,qh, 0)}} \rightarrow {e^{i\pi/4}}\,
{{\Bigl({\eta_{13}}{\eta_{24}}\Bigr)^{1\over 4}}
\over{(1+\sqrt{x})^{1/2}}}\,\,
\Biggl({{1+\sqrt{x}}\over \sqrt{x}}{\Psi_{(13)(24)}}\,\,+
\,\,{{1-x}\over\sqrt{x}}\,{\Psi_{(12)(34)}}\Biggl)
\eqn\psiidtranstwo
$$
or,
$$
{\Psi^{(4\,qh, 0)}} \rightarrow {e^{i\pi/4}}\,
{\Bigl({\eta_{13}}{\eta_{24}}\Bigr)^{1\over 4}}\,
\Biggl({{(1+\sqrt{x})^{1/2}}\over \sqrt{x}}{\Psi_{(13)(24)}}\,\,+
\,\,{{(1-\sqrt{x})^{1/2}}\over\sqrt{x}}\,
\sqrt{1-x}\,{\Psi_{(12)(34)}}\Biggl)~.
\eqn\psiidtransthree
$$
That this is, in fact, equal to
$$
{\Psi^{(4\,qh, 0)}}\rightarrow\,{{e^{i\pi/4}}\over\sqrt{2}}
\Biggl({\Psi^{(4\,qh, 0)}}+{\Psi^{(4\,qh, 1/2)}}\Biggr)
\eqn\psiidtransfour
$$
can be seen with the use of the identity:
$$
{\Bigl(1-\sqrt{1-{x}}\Bigr)^{1\over 2}}\,\pm\,
{\Bigl(1-\sqrt{1-{x}}\Bigr)^{1\over 2}}=
\pm\sqrt{2}{\Bigl(1\pm\sqrt{x}\Bigr)^{1\over 2}}
\eqn\psiidtransid
$$
The correspondence between $SO(4)$ rotations
and the action of braiding particles
can be shown, in a similar fashion,
for the other possible braids.

To extend these considerations to
the general case of $2n$ quasiholes,
imagine bringing four of the quasihole operators
close together in the appropriate conformal block.
The braiding of one of these four quasiholes around another
is governed by the operator product
expansion, and therefore is generated  by
the transformations we found above in the four quasihole
case.  This is true for any group of four
quasiholes, which allows us to make rotations in any chosen plane
of configuration space.  
In this way we construct a correspondence
between the elementary exchanges which generate
the braid group and the rotations which generate $SO(2n)$.

\chapter{Black Hole Analogy}

A large number of quasiholes with nearby centers carve out a
macroscopic, geometric hole in a Hall droplet.  Conversely, one constructs  
wave functions describing an
annular quantum Hall fluid by 
notionally inserting an appropriate set of quasiholes.  It should be
noted that the quasiholes in this construction lie outside the bulk
fluid; in this sense they resemble the image charges of electrostatics.
To create the deficit appropriate to an area $A$ in the Pfaffian
state, one must insert $n_{\rm A} = {eB\over \pi}  A$ quasiholes.  

According to the foregoing analysis, there is a degeneracy
$2^{{n_{\rm A}\over 2}  -1}$ associated to this many quasiholes at fixed
positions. Thus there is an entropy
$$
S~=~ ({n_{\rm A}\over 2} -1) \ln 2  ~\approx ~ {\pi \ln 2 \over 2eB} A
\eqn\entropy
$$
proportional to the area of the geometric hole.  
This degeneracy is quite independent of the possibility of edge
excitations, which correspond to different geometrical arrangements of
the image quasiholes. 

One notices in \entropy\ 
at least a metaphorical resemblance to the entropy of an
extremal -- zero temperature -- 
black hole.  A notable contrast is that the entropy here is
proportional to the total area, as opposed to the surface area ({\it
i.e}. the perimeter) of the hole.  As we have discussed, the 
states under consideration generally
differ only with respect to subtle high-order
correlations among the electrons.  In principle, however, all the
information is in the exterior wave function -- these black hole
analogues have
plenty of hair.

Is there an analogue of Hawking radiation?  There is, although it
lacks the universality of its gravitational cousin.  One might imagine
that at the inner edge of the Hall annulus there could be a potential
gradient,
so that it would be energetically favorable for the droplet to slowly
contract.  At finite temperature, this tendency would be countered by
the loss of entropy associated with decrease of $A$.  The free energy
would balance at a temperature depending on $A$ (and the potential).
If the system were not closed, but allowed to radiate, there could be
an evolution mimicking that expected of black holes.  
Since one has for the analogue system an adequate, completely
conventional quantum-mechanical description -- though the states are
certainly unusual and exotic in detail --  the question of information
loss does not seriously arise.

For another analogy between quantum Hall edge states and black holes,
which seems quite different, see [\bala ]

\chapter{Discussion}

%%%%%%%%%%%%%%%%%%%%%%

\section{Connection to $(3,3,1)$ State} 

The so-called $(3,3,1)$ state is another incompressible quantum Hall
state at $\nu = 1/2$.  It can arise when one has two distinguishable
classes of electrons, such as in a double layer system or when both
spin up and spin down are relevant.  We will not review the standard
lore on this state here.

Recently in [\ho] T.L. Ho pointed out that the $(3,3,1)$ state
in double-layer quantum Hall systems
is a paired state that may be continuously
deformed into the Pfaffian state.  At a first glance,
this result seems to lead to a contradiction, because there are
integers intrinsically attached to these states which distinguish
between them.    
Specifically, the $(3,3,1)$ state is believed to
have degeneracy $8$ on the torus
while the Pfaffian
has degeneracy $6$.  Closely related to this, the statistics of
quasiparticles in the $(3,3,1)$ state is 
supposed to be Abelian, while the
Pfaffian state supports quasiparticles with non-Abelian statistics.
Upon closer scrutiny, we will see that there
is in fact no contradiction and, in particular, that
non-Abelian statistics might be observed in a refined treatment of  
systems crudely described by the $(3,3,1)$ state
(see also [\flt]).

A central point of Ho's paper 
is that the classic $(3,3,1)$ droplet ground
state wavefunction may be
written in the form:
$${\Psi_{(3,3,1)}} = {\rm Pf} \biggl(
{{{u_i}{v_j} + {u_j}{v_i}}\over{{z_i}-{z_j}} }
\biggr)
\,\,\prod {({z_i}-{z_j})^2}\eqn\psitto$$
where $u$ and $v$ are up- and down-pseudospin spinors.
(Ho's notation is rather different.)  The equivalence of \psitto\ 
to the conventional form of the $(3,3,1)$ wavefunction is a form of
Cauchy's identity.

If we replace $v$ in \psitto\
by ${\tilde v}$ and let ${\tilde v}$
continuously interpolate between $v$
and $u$, then we will have a sequence of states
which interpolate between the $(3,3,1)$
and Pfaffian states. The $(3,3,1)$ state
has half-flux quasiholes, which
can be placed in either layer. A state with one
quasiparticle in the upper layer at $\eta_1$
and one quasihole in the lower layer
at $\eta_2$ would look like:
$${\Psi_{(3,3,1)}^{2\,qh}} = {\rm Pf} \biggl(
{{{u_i}{v_j}({z_i}-{\eta_1})({z_j}-{\eta_2}) +
{u_j}{v_i}({z_j}-{\eta_1})({z_i}-{\eta_2})}\over{{z_i}-{z_j}} }
\biggr)
\,\,\prod {({z_i}-{z_j})^2}~.\eqn\psittotqh$$
However,
$$\eqalign{
{u_i}{v_j}&({z_i}-{\eta_1})({z_j}-{\eta_2}) +
{u_j}{v_i}({z_j}-{\eta_1})({z_i}-{\eta_2}) =\cr
&({u_i}{v_j} + {u_j}{v_i})\Bigl(({z_i}-{\eta_1})({z_j}-{\eta_2})
+ ({z_j}-{\eta_1})({z_i}-{\eta_2})\Bigr)
+({z_i}-{z_j})({\eta_1}-{\eta_2})({u_i}{v_j} - {u_j}{v_i})
\cr ~.}\eqn\psittoid$$
If we substitute this into \psittotqh\ all 
terms will vanish upon antisymmetrization
except for those with no spin-singlet pairs or a single
spin-singlet pair:
$$\eqalign{
{\Psi_{(3,3,1)}^{2\,qh}} &= {\rm Pf} \biggl(
{{ ({u_i}{v_j} + {u_j}{v_i})\Bigl(({z_i}-{\eta_1})({z_j}-{\eta_2})
+ ({z_j}-{\eta_1})({z_i}-{\eta_2})\Bigr)
}\over{{z_i}-{z_j}} }
\biggr)
\,\,\prod {({z_i}-{z_j})^2}\cr
&\qquad + {\cal A}\biggl(
({\eta_1}-{\eta_2})({u_1}{v_2} - {u_2}{v_1})\,\,
{{ ({u_3}{v_4} + {u_4}{v_3})\,\,\Bigl(({z_3}-{\eta_1})({z_4}-{\eta_2})
+ {({z_4}-{\eta_1})({z_3}-{\eta_2})\Bigr)
}\over{{z_3}-{z_4}} } }\cr
&\qquad{ {{ ({u_5}{v_6} + {u_6}{v_5})
\Bigl(({z_5}-{\eta_1})({z_6}-{\eta_2})
+ ({z_6}-{\eta_1})({z_5}-{\eta_2})\Bigr)
}\over{{z_5}-{z_6}} } }\,\,\ldots
\biggr)\prod {({z_i}-{z_j})^2}\cr}
\eqn\psittotqps$$
By arguments similar to those below equation \psichisub , the
$2n$-quasihole state can have $k=0,1,\ldots,n$ spin-singlet
pairs. These states form the representations
of $SU(n)$ consisting of tensors with
$k$ antisymmetric upper indices
into which the spinor representation of $SO(2n)$
decomposes.

Ho considered a model Hamiltonian for which spin-singlet
pairs are energetically unfavorable, specifically:
$$\eqalign{
H &= \,{\sum_{i\neq j}}\, {P_{ij;S=0}}
\,\Bigl({V_s^0}\,\delta({z_i}-{z_j})
+{V_s^2}\,\delta''({z_i}-{z_j})\Bigr)\cr
&\qquad +\,\, 
{V_t}\,{\sum_{i>j}}\,
\Bigl( {P_{ij;S=1,{S_z}=1}}\,+\,{P_{ij;S=1,{S_z}=-1}} \Bigr)
\,\delta'({z_i}-{z_j})~.\cr}
\eqn\hoham$$
Here
${P_{ij;S,{S_z}}}$ projects ${{\bf S}_i}+{{\bf S}_j}$
on a state of spin $(S,{S_z})$.  
The multi-quasihole states of the $(3,3,1)$ state
are in one-to-one correspondence with those of
the Pfaffian, and it is no surprise that the
two states can be adiabatically connected. The
${u_i}{v_j} + {u_j}{v_i}$ factors are a trivial
modification of the Pfaffian, so the non-Abelian
statistics which we found earlier is exhibited
by the quasiholes of this form of the $(3,3,1)$ as well.

Furthermore {\it for such a Hamiltonian\/} 
the $(3,3,1)$ state is $6$-fold degenerate
on the torus.  Indeed, the
Pfaffian part of the wavefunction takes the following form
on the torus:
$$\eqalign{{\rm Pf} \biggl(
{{({u_i}{v_j} + {u_j}{v_i})\,{\theta_a}({z_i}-{z_j})}\over
{{\theta_1}({z_i}-{z_j})} }
\biggr) &= {\cal A} \Biggl({ {\prod{{\theta_1}({z_{2i}}-{z_{2j}})}
\,\prod{{\theta_1}({z_{2i-1}}-{z_{2j-1}})} }\over
\prod{{\theta_1}({z_{2i}}-{z_{2j-1}})} }\,\times\cr
&\qquad\qquad\,{\theta_a}
\Bigl(\sum\,{z_{2i}}-{z_{2i-1}}\Bigr)\,
\prod{u_{2i}}{v_{2i-1}}
\Biggr)\cr}
\eqn\psisfold$$
where $a=2,3,4$. This 3-fold degeneracy, together
with the 2-fold center-of-mass degeneracy leads
to an overall 6-fold degeneracy, as in the Pfaffian
state. In other words, for this Hamiltonian,
the degeneracy and quasihole statistics
are the same for the $(3,3,1)$ and for Pfaffian
states -- and, of course, for all the states that
interpolate between them.

On the other hand, if we had taken the Hamiltonian:
$$\eqalign{H &= {\sum_{i\neq j}}\, {P_{i,{S_z}=1/2}}\,{P_{j,{S_z}=-1/2}}\,
\delta({z_i}-{z_j})\cr
&\qquad + 
{\sum_{i>j}} \Bigl({P_{i,{S_z}=1/2}}\,{P_{j,{S_z}=1/2}}+
{P_{i,{S_z}=-1/2}}\,{P_{j,{S_z}=-1/2}}\Bigr)
\,\delta''({z_i}-{z_j})\cr}
\eqn\hamtto$$
(${P_{i,{S_z}}}$ projects ${{\bf S}_i}$ on a
state with $z$-component ${S_z}$) then
spin singlet-pairs would not be energetically
unfavorable, and states containing them would be part of
the degenerate quasihole spectrum.

Because of this distinction, it is appropriate to write
$(3,3,1)_p$ when the projected (paired, Pfaffian) universality class
is intended, and $(3,3,1)_c$ when the classic universality class is
intended.  The standard model wave function for the droplet ground
state is the same in either case, but the excitation spectra and the
ground state degeneracy on a torus differ, as we have seen.

The additional type of excitation occurring in a system with
the Hamiltonian \hamtto\ leads to an increase
in the degeneracy on the torus,  because the
degenerate ground states are obtained by creating a pair of
excitations, winding one of them around a non-trivial
cycle on the torus, and annihilating them.
The extra two states, which lead to an 8-fold
degeneracy, have \psisfold\ replaced by:
$${\cal A} \Biggl({ {\prod{{\theta_1}({z_{2i}}-{z_{2j}})}
\,\prod{{\theta_1}({z_{2i-1}}-{z_{2j-1}})} }\over
\prod{{\theta_1}({z_{2i}}-{z_{2j-1}})} }
\,\,{\theta_1}\Bigl(\sum{z_{2i}}-{z_{2i-1}}\Bigr)\,
\prod{u_{2i}}{v_{2i-1}}
\Biggr)~.\eqn\psiefold
$$

In general the terms in both types of Hamiltonians will be present at
some level. The state with less degeneracy, {\it i.e}. the Pfaffian,
will in principle give the finer description.  Of course if the terms
that exact an energetic price for the singlets are small, as they
might well be for a double-layer system (since they involve effects
depending on coherence between the layers),
then the splittings will be small.
As ${V_s^2}\rightarrow 0$ in \hoham, these
splittings go to zero. For ${V_s^2}< 0$, spin-singlets
are favored. For $|{V_s^2}|$ large, this will lead
to the breakdown of the state because the
creation of quasiparticle-quasihole pairs with
associated spin-singlets as in the second term in
\psittotqps\ will be energetically favored.
For $|{V_s^2}|$ small, however, it is possible
that the $(3,3,1)$ ground state is stable
while the quasihole states with spin-singlet pairs
have negative energy.

\section{Connection to Merons}

%%%%%%%%%%%%%%%%%%%%

If we allow for spin, and include the effect of Coulomb interactions,
then in the low Zeeman energy limit, the Laughlin quasiholes deform
into Skyrmions [\sondhi].  In the same circumstances, we expect that the
half-flux quasiholes in the paired state become {\it merons}
[\bloom,\gross].  

To appreciate this, let us briefly recall 
the essence of the Skyrmion construction [\bloom].
To construct a Skyrmion at the origin, one multiplies the ground state
wave function by a spin-dependent factor as follows:
$$
{\Psi_{\rm skyr.}}(z_j) ~=~ {\Psi_0}\, {\prod_i}
{ {z_i} \choose \lambda }
\eqn\skyrwf
$$
The spins of the electrons point up far from the
origin, but point down at the origin. The charge
deficit, which can be inferred from the form of
the wavefunction at infinity, is spread over
a region of radius $\sim\lambda$.
Wavefunctions with skyrmions at ${\eta_1},{\eta_2},\dots,{\eta_n}$
have the spinor factor in \skyrwf\
replaced by [\multiskyr]
$$
{\prod_i}\Biggl( {\prod_\alpha}({z_i}-{\eta_\alpha})\,
{1 \choose {\lambda {\sum_\alpha} {1\over{{z_i}-{\eta_\alpha}}}}}
\Biggr)
\eqn\manyskyr
$$

Let us write our Pfaffian factor in the
form
$$
{\rm Pf} \biggl( {{{h_i}{k_j} + {h_j}{k_i}}\over{{z_i}-{z_j}} }\biggr)~,
\eqn\genpf
$$
where in the polarized 
ground state the spinor-valued 
functions $h$ and $k$ simply point up everywhere.  Now we can apply
the `multi-skyrmion' construction {\it separately and independently 
on the h and k factors}:
$$\eqalign{
{h_i}&={\prod_i}\Biggl(\, {\prod_{\alpha\in\Lambda}}({z_i}-{\eta_\alpha})\,
{1 \choose {\lambda {\sum_\alpha} {1\over{{z_i}-{\eta_\alpha}}}}}
\Biggr)\cr
{k_i}&={\prod_i}\Biggl(\, {\prod_{\alpha\in{\bar\Lambda}}}({z_i}-{\eta_\alpha})
\,
{1 \choose {\lambda {\sum_\alpha} {1\over{{z_i}-{\eta_\alpha}}}}}
\Biggr)\cr}
\eqn\multimeron
$$
where $\Lambda$ and ${\bar \Lambda}$ are the two macrogroupings.
In this way half-skyrmions,
or merons, appear in a natural and
canonical fashion.

This construction gives a new perspective on the significance of the
macro-groupings.  Each macro-grouping of quasiholes contains merons of
a single type; only by combining two of opposite types 
do we obtain a Skyrmion.

In this rather telegraphic indication, for the sake of brevity we 
have ignored some subtle but important 
refinements of the construction, necessary if we are
to keep the textured spin
directions associated with definite spatial positions and to patch
together multi-skyrmions (or multi-merons) while keeping the
wavefunction holomorphic and properly correlated.  A fully
satisfactory version can be obtained by following closely the procedures
previously described in [\ourtextures ].

\section{Possible Experimental Consequences}

We will only attempt a few broad remarks concerning the possible
experimental consequences of the structures we have been discussing
here.  More work, of a different character, will be necessary to give
realistic quantitative assessments of the possibilities.

We have already mentioned the interest of probing annular geometries
for analogues of the Hawking process, and the possibility of
low-temperature phase transitions. 

We have, for a large number of well-separated quasiholes, identified
many quantum states
that differ only in subtle high-order correlations.  Insofar as
thermal equilibrium is established among these states, there will be
obvious consequences for thermodynamic quantities such as the specific
heat.  On the other hand the processes tending to establish
equilibrium appear to involve large-scale rearrangements and may be
highly suppressed.  The most fundamental such processes involve
braiding of the quasiholes; but in realistic circumstances these
quasiholes will tend to be pinned by impurities.  We therefore
anticipate glassy behavior, {\it i.e}. that the effective spin-like degrees
of freedom we have identified form a spin glass.  

Finally let us note that in the paired, `Pfaffian' version of the $(3,3,1)$ 
state, with singlets projected out, the axis of pairing forms a macroscopic
quantum variable.  It couples to a potential difference between the
layers, and the possibility of Josephson-like effects arises
(Compare [\wzj,\bloom,\ho]).  These possibilities
are currently under investigation.

\bigskip
\bigskip

{\bf Acknowledgment}: We would like to thank G. Baskaran, M. Flohr, V.
Gurarie, F.D.M. Haldane, S. Sondhi, and K. Yang for helpful discussions.

%%%%%%%%%%%%%%%%%%%%

\endpage
\refout

\end